\documentclass[12pt,preprint]{aastex}

\usepackage{emulateapj5}
\usepackage{apjfonts}

\renewcommand\tablecaption[1]{\centerline{\sc #1}\vskip 6pt}
\renewcommand\colhead[1]{\normalsize{#1}}
\renewenvironment{deluxetable}{\begin{tabular}}{\end{tabular}}

\renewcommand\plotone[1]{\epsffile{#1}}
\renewenvironment{figure}{\epsfxsize 3.5in}{\vskip 12pt}

\newcommand{\cago}{$^{12}{\rm C}(\alpha,\gamma)^{16}{\rm O}$\ }

\shorttitle{ASTEROSEISMOLOGY AND THE \cago RATE}
\shortauthors{TRAVIS S. METCALFE}

\begin{document}

\title{White Dwarf Asteroseismology and the \cago Rate}

\author{Travis S. Metcalfe}

\affil{Harvard-Smithsonian Center for Astrophysics,
       60 Garden Street, Cambridge, MA 02138}

\email{tmetcalfe@cfa.harvard.edu}

\submitted{Accepted for publication in ApJ Letters}

\begin{abstract}
Due to a new global analysis method, it is now possible to measure the
internal composition of pulsating white dwarf stars, even with relatively
simple theoretical models. The precise internal mixture of carbon and
oxygen is the largest single source of uncertainty in ages derived from
white dwarf cosmochronometry, and contains information about the rate of
the astrophysically important, but experimentally uncertain, \cago nuclear
reaction. Recent determinations of the internal composition and structure
of two helium-atmosphere variable (DBV) white dwarf stars, GD~358 and
CBS~114, initially led to conflicting implied rates for the \cago
reaction. If both stars were formed through single-star evolution, then
the initial analyses of their pulsation frequencies must have differed in
some systematic way. I present improved fits to the two sets of pulsation
data, resolving the tension between the initial results and leading to a
value for the \cago reaction rate that is consistent with recent
laboratory measurements.
\end{abstract}


\section{Introduction}

As a white dwarf star cools over time, from a hot planetary nebula nucleus
down to the coolest temperatures possible given the finite age of the
universe, there are three narrow ranges of surface temperature where they
may become pulsationally excited. The particular interval where a given
white dwarf will pulsate is determined by its spectral type. The PG~1159
stars or variable DO white dwarfs are the hottest class, and require
detailed evolutionary models for an accurate pulsational analysis. The
cooler helium-atmosphere (DB) and hydrogen-atmosphere (DA) variables are
both easier to model, since relatively simple polytropic models may be
used to approximate their evolution quite accurately \citep{woo90}. The DB
variables (DBVs) are structurally the simplest, since they have a single
surface layer of helium on top of a presumed carbon/oxygen (C/O) core,
while DA variables (DAVs) have an additional surface layer of hydrogen
above this helium mantle.

Although our theoretical models of variable white dwarfs are relatively
simple, they have been able to match the observed pulsation periods of
these stars with a degree of accuracy that---with the exception of the
Sun---is unsurpassed in the field of stellar seismology. In part, this
achievement has been made possible by the simplifying physical
circumstances: the absence of nuclear fusion, the extremely high surface
gravity ($\log g\sim8$), slow rotation ($P_{\rm rot}\sim1$ d), and
negligible magnetic fields. However, the most recent improvements in the
agreement between models and observations have been driven more by global
explorations of the defining parameters of existing models, rather than by
fundamental changes to the models themselves.

By combining improved models with objective global optimization methods,
we may eventually obtain physically complete and fully self-consistent
theoretical models of white dwarf stars. Until then, we can rely on the
{\it relative} quality of the match between our models and observations as
a proxy for the {\it absolute} optimization that would be possible if our
models were physically complete descriptions of real white dwarfs. This
may lead to systematic errors in the values of some of our parameters,
depending on the nature of the incompleteness, but it will allow us to
proceed and to develop useful algorithms for diagnosing the limiting
assumptions in our models.

Recently, \citet{mc03} have given a detailed description of a method
developed to match simple models of DBV white dwarfs to the available
observations. The method itself is perfectly general, and may easily be
extended to other types of pulsating stars. In the context of DBV white
dwarfs it has been used, among other things, to infer the internal mixture
of carbon and oxygen---the most important source of uncertainty in age
estimates from white dwarf cosmochronometry---and, by extension, the rate
of the key \cago reaction. The initial applications of this method to two
white dwarfs, first to the brightest DBV star, GD~358 \citep{msw02}, and
then to the faintest member of the class, CBS~114 \citep{hmw02}, yielded
apparently conflicting results. Here, I will demonstrate that the apparent
conflict arose from systematic differences in how the two data sets were
analyzed. When the optimum models are determined using identical criteria,
both stars yield internal compositions that imply rates for the \cago
reaction that are consistent with one another, and which both agree with
recent extrapolations from high-energy laboratory measurements.

\section{Initial Results}

The global optimization method for DBV white dwarf models, based on the
publicly available genetic algorithm {\tt PIKAIA} \citep{cha95}, was
developed by \citet{mnw00} and later extended by \citet[hereafter
MWC]{mwc01} to allow five adjustable parameters. The method attempts to
minimize the differences between the observed and calculated periods and
period spacings for models with effective temperatures ($T_{\rm eff}$)
between 20,000 and 30,000 K, total stellar masses ($M_*$) between 0.45 and
0.95 $M_{\odot}$, helium layer masses with $-\log(M_{\rm He}/M_*)$ between
2.0 and $\sim$7.0, and an internal C/O profile with a constant oxygen mass
fraction ($X_{\rm O}$) out to some fractional mass ($q$) where it then
decreases linearly in mass to zero oxygen at $0.95~m/M_*$. The broad range
of the search is limited only by observational constraints and by the
physics of the models.

The observational data for GD~358 came from an extremely successful
multi-site campaign in 1990 by the Whole Earth Telescope collaboration
\citep[WET;][]{nat90}. The results and analysis of these observations were
reported by \citet{win94} and \citet[hereafter BW]{bw94} who conclusively
identified a series of 11 consecutive radial overtones ($k$=8-18) of
non-radial g-mode pulsations with the same spherical degree ($\ell$=1).
Using this rich data set, BW attempted to match the periods themselves
($P_k$) as well as the spacings between consecutive modes ($\Delta P
\equiv P_{k+1}-P_k$). When confined to search the same range of parameters
considered in this initial study, MWC found the same optimal set of model
parameters as BW.  But the global search revealed a significantly better
model outside the range of the initial search, with optimal parameters
$T_{\rm eff}=22,600$ K, $M_*=0.650~M_{\odot}$, $\log(M_{\rm
He}/M_*)=-2.74$, $X_{\rm O}=0.84$, $q=0.49~m/M_*$, and root-mean-square
(rms) residuals $\sigma_P=1.28$ s, and $\sigma_{\Delta P}=1.42$ s.

The central oxygen mass fraction in white dwarf models is primarily
determined by the rate of the \cago reaction. The extent and efficiency of
internal mixing during the red giant phase is a secondary factor (see \S
\ref{discussionsec}). This allowed \citet{msw02} to adjust the \cago rate
in evolutionary models of the internal chemical profiles until they
produced the optimal value of $X_{\rm O}$ for GD~358. The implied reaction 
rate was
$S_{300}=370\pm40$ keV barns, significantly higher than most
extrapolations from high-energy laboratory measurements. The authors noted
that the application of this method to additional DBV white dwarfs could
provide independent measurements of this important nuclear reaction rate.

Although CBS~114 is the faintest known DBV, single-site observations from
SAAO in 2001 \citep{hmw02}, combined with a reanalysis of the discovery
data of \citet{wc88}, revealed 7 stable pulsation modes with periods
between about 400-650 seconds. The mean period spacing was consistent with
non-radial g-mode pulsations of spherical degree $\ell$=1, and a
comparison with models identified the radial overtones as $k$=8,9 and
11-15, with the $k$=10 mode undetected. The global analysis of these
pulsation data had to be restricted to comparing the observed and
calculated periods---neglecting the period spacings because of the
undetected mode that interrupts the sequence of otherwise consecutive
radial overtones. The optimal values for the five model parameters were
$T_{\rm eff}=21,000$ K, $M_*=0.730~M_{\odot}$, $\log(M_{\rm
He}/M_*)=-6.66$, $X_{\rm O}=0.61$, $q=0.51~m/M_*$, and the rms period
residuals were $\sigma_P=0.43$ s. This fit matched the observed periods so
closely that it also reproduced the observed deviations from the mean
period spacing (an alternate period spacing criterion) almost exactly
\citep[see][Fig.~9c]{hmw02}.

The higher stellar mass for CBS~114 relative to GD~358 should naturally
lead to a {\it slightly lower} central oxygen mass fraction, since the
\cago reaction is slightly less efficient at higher densities. However,
the internal chemical profile models with the appropriate mass required a
rate for the \cago reaction of only $S_{300}\approx180$ keV barns to
reproduce the optimal central oxygen mass fraction. This is very close to
the rate derived from recent laboratory measurements \citep[$S_{300}=
165\pm50$ keV barns;][]{kun01}.

The conflicting results for GD~358 and CBS~114 led \citet{hmw02} to
speculate that either some source of systematic uncertainty in the models
was affecting the analysis of the two stars in different ways, or that
they may have different evolutionary origins. The easiest way to test the
second idea would be to apply the method to an additional DBV star, and
see if the internal composition agreed with either of the other two.
Although the WET recently conducted a multi-site campaign on the DBV star
PG~1456+103, the analysis of those data are still in progress. Thus, I
have attempted to follow up on the first possibility, and identify
possible sources of systematic error.

One obvious systematic problem with both of the fits, as noted in each of
the initial analyses, was the disagreement between the optimal masses and
temperatures and the values of these parameters from the spectroscopic
analysis of \citet{bea99}. Both global fits produced masses that were
systematically high, and temperatures that were systematically low. Some
discrepancy was expected, since the two methods assumed slightly different
values for the mixing-length/pressure scale height ratio for atmospheric
convection. In addition, the simple polytropic models used for the global
fits included static envelopes containing 5\% of the total stellar mass.
\citet{msw02} demonstrated that by reducing the fractional mass of these
static envelopes to only 2\%, the mass and temperature of the optimal
model for GD~358 were in closer agreement with the spectroscopic values.
In any case, neither of these two problems should be responsible for the
differences between the results for GD~358 and CBS~114, since both
problems were common to the two analyses.

\begin{figure}
\plotone{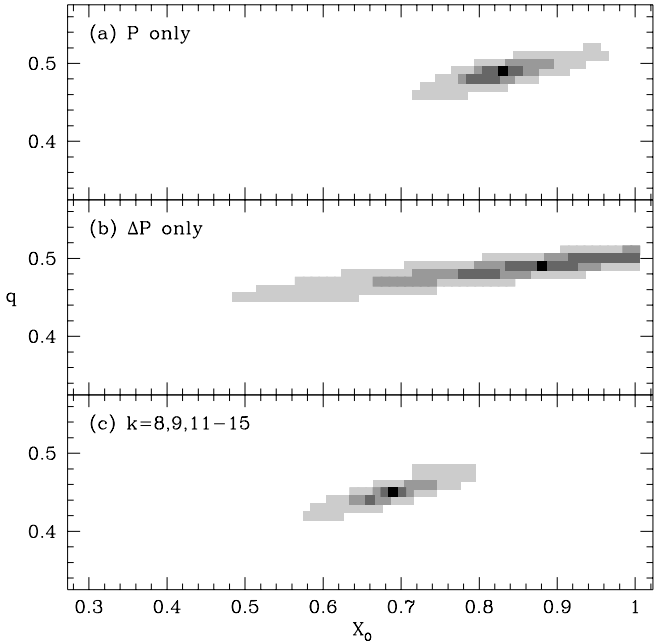}
\figcaption[f1.eps]{The optimal combinations of $X_{\rm O}$ and $q$ for
GD~358 (black) and the distribution of parameter values that yield rms
residuals within 1, 3, and 10 times the observational uncertainty
(progressively lighter shades of grey) when the mass, temperature and
helium layer mass are fixed at their optimal values from MWC and when the
fitting criterion is based on matching (a) the pulsation periods only, (b)  
the period spacings only, and (c) the periods of only those pulsation
modes observed in CBS~114.\label{fig1}}
\end{figure}

Upon reflection, there were two things that {\it did} differ between the
analyses of the two stars: (1) the fit to GD~358 used the periods {\it
and} the period spacings to judge the quality of the match, and (2) there
were a larger number of higher radial overtone modes in the case of
GD~358. To evaluate whether or not these subtleties could be responsible
for the differences between the derived values of the central oxygen mass
fraction, we can look at a small grid of models with various values of
$X_{\rm O}$ and $q$ keeping the mass, temperature, and helium layer mass
fixed at their optimal values. The effect of these two differences on the
determination of the central oxygen mass fraction for GD~358 are
illustrated in Figure~\ref{fig1}. Each panel shows the optimal combination
of $X_{\rm O}$ and $q$ for a given fitting criterion in black, along with
the ranges of model parameters leading to rms residuals that differ from
those of the optimal combination by 1, 3, and 10 times the observational
uncertainty ($\sigma_{\rm obs} \sim 0.03$ s), shown in progressively
lighter shades of grey. The difference between the distributions in the
top two panels demonstrates that when the period spacings are used in
addition to the periods themselves for fitting, the effect is to bias the
central oxygen mass fraction toward higher values and to increase the
uncertainty in the optimal value of $X_{\rm O}$. Comparing the top and
bottom panels, it is clear that attempts to match only those radial
overtones observed in CBS~114 may lead to systematically lower values for
$X_{\rm O}$. In principle, these two results would lead us to expect that
the initial determination of $X_{\rm O}$ for GD~358 may have overestimated
the actual value due to the use of period spacings for fitting, and the
value for CBS~114 may have been an underestimate since there were fewer
modes of higher radial overtone. In practice, this exercise cannot provide
{\it quantitative} information about the expected shifts, since the three
model parameters that were fixed for Figure~\ref{fig1} will undoubtedly
shift slightly in a full re-optimization.

\section{New Model-Fitting}

Motivated by the qualitative evidence from Figure~\ref{fig1}, I performed
several new fits using the global model-fitting procedure of MWC. Since I
also wanted to reconcile the discrepancies between the masses and
temperatures inferred from asteroseismology and spectroscopy, the new fits
used the so-called ``ML2/$\alpha$=1.25'' prescription for convection
adopted for the spectroscopic fits by \citet{bea99} rather than the
``ML3'' prescription used by MWC. Both prescriptions follow the same
mixing-length theory of \citet{bc71}, but assign different values (1.25
and 2.0 respectively) to the mixing-length/pressure scale height ratio.
Each of the new fits use only the periods (and {\it not} the period
spacings) to determine the optimal model, which also makes them directly 
comparable to the alternative model fit of \citet{fb02}.

The values of the globally optimal model parameters for the new fits are
listed in Table~\ref{tab1}, along with the rms differences between the
observed and calculated periods ($\sigma_P$)\footnotemark[1]. 
 \footnotetext[1]{An additional fit for GD~358, using ML2/$\alpha$=1.25
 convection and {\it both} P and $\Delta$P led to the optimal parameters
 (22,900 K, 0.645 M$_{\odot}$, $-$2.69, 0.63, 0.46), yielding $\sigma_P$=
 1.32 and $\sigma_{\Delta P}$=1.50. This suggests that the lower value for
 $X_{\rm O}$ is primarily due to the change of convective prescription.}
Note that the period 
residuals for the ``GD~358 [11]'' fit are considerably smaller than in the
fit of MWC (which had $\sigma_P=1.28$ s). The period residuals for CBS~114
also improved slightly, compared to the fit of \citet[which had
$\sigma_P=0.43$ s]{hmw02}. Small grids of models with various combinations
of $X_{\rm O}$ and $q$, keeping the other parameters fixed, revealed that
the formal 1$\sigma$ statistical uncertainties on the central oxygen mass
fraction for GD~358 and CBS~114 are near $\pm$0.01 in both cases, but I
adopt an uncertainty twice this size.

Using the mass and central oxygen mass fraction to derive a measurement of
the \cago reaction rate is model-dependent. To facilitate comparison with
previous results, I relied on the internal chemical profiles of Metcalfe,
Salaris, \& 

\begin{figure}
\plotone{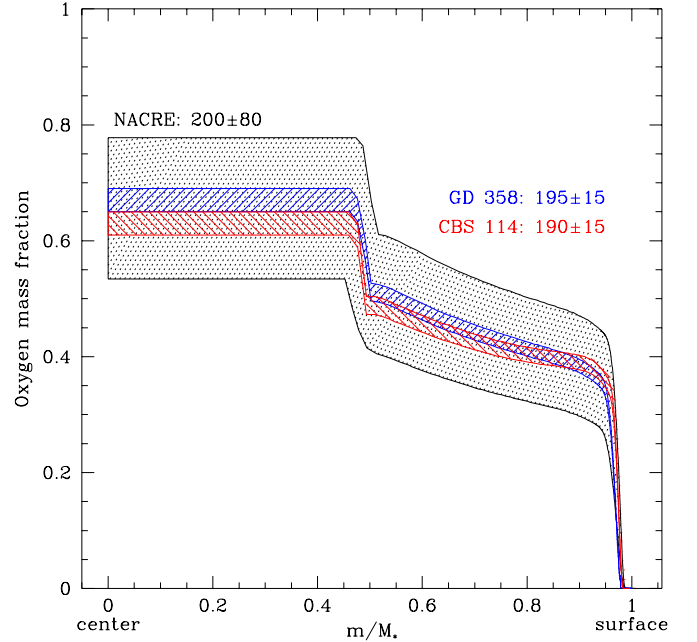}
\figcaption[f2.eps]{The $\pm1\sigma$ range of theoretical internal chemical
profiles for a 0.65 $M_{\odot}$ white dwarf model using the NACRE rate for
the \cago reaction \citep[shaded area]{ang99}, along with profiles scaled
to the optimal central oxygen mass fraction derived for GD~358 (upper
hashed area) and CBS~114 (lower hashed area) with the corresponding values
for the \cago rate.\label{fig2}}
\end{figure}

\noindent 
Winget (2002) and M.~Salaris (private communication). Interpolating
between these models, the \cago reaction rate that was required to match
the optimal values of the mass and central oxygen mass fraction for GD~358
and CBS~114 was $S_{300}=195\pm15$ keV barns and $S_{300}=190\pm15$ keV
barns respectively. Both measurements are in excellent agreement with the
rate suggested by the NACRE collaboration \citep[$S_{300}=200 \pm 80
$;][]{ang99}.  This remarkable
agreement is illustrated in Figure~\ref{fig2}, which shows the ranges of
internal chemical profiles corresponding to the NACRE rate for a 0.65
$M_{\odot}$ white dwarf model \citep[from][]{msw02} along with profiles
scaled to match the optimal values of the central oxygen mass fraction for
GD~358 and CBS~114. Note that the \cago rates derived from the two stars
are statistically indistinguishable, but the chemical profiles themselves
do not overlap at the center because the models have different masses.
Both measurements are also consistent with the more recent laboratory
value derived by \citet[$S_{300}=165\pm50$]{kun01}.

\begin{table*}
\begin{center}
\vspace*{12pt}
\tablenum{1}
\centerline{\sc Table \ref{tab1}}
\tablecaption{New model-fits for two DBV stars\label{tab1}}
\begin{deluxetable}{lrrr}
\hline\hline
\colhead{Parameter}    &
\colhead{GD~358 [11]}  &
\colhead{GD~358 [7]}   &
\colhead{CBS~114}      
\\ \hline
$T_{\rm eff}~(K)$\dotfill   & 22,900       & 22,100     & 20,300  \\
$M_*~(M_{\odot})$\dotfill   & 0.660        & 0.665      & 0.750   \\
$\log[M_{\rm He}/M_*]\ldots$& $-$2.00      & $-$2.79    & $-$6.82 \\
$X_{\rm O}$\dotfill         & 0.67         & 0.76       & 0.63    \\
$q~(m/M_*)$\dotfill         & 0.48         & 0.47       & 0.51    \\
$\sigma_P~(s)$\dotfill      & 1.05         & 0.52       & 0.41    \\
\hline\hline
\end{deluxetable}
\end{center}
\vspace*{-24pt}
\end{table*}

\section{Discussion \label{discussionsec}}

A re-analysis of the pulsation data for GD~358 and CBS~114, with
ML2/$\alpha$=1.25 convection and using only the
periods to determine the optimal models, yielded improved fits leading to
measurements of the \cago reaction rate that are both consistent with each
other and with laboratory measurements. Since a larger number of pulsation
modes are present in GD~358 to constrain the fit, it should be considered
the more reliable measurement. The influence of having fewer modes and
lower radial overtones on the derived central oxygen mass fraction is
ambiguous: for a given mass, temperature, and helium layer mass, it may
lead to an underestimate---but when all parameters are allowed to
re-adjust (as in the ``GD~358 [7]'' fit in Table \ref{tab1}) it may lead
to an overestimate. This underscores the importance of a global analysis
when trying to determine the optimal model.

The derived central oxygen mass fractions for these two DBVs are now
perfectly consistent with each other---with a slightly lower value for the
more massive white dwarf, as expected. Mapping the internal chemical
composition into a rate for the \cago reaction is model-dependent, as
discussed recently by \citet{str03}, because the adopted mixing scheme and
numerical treatment of breathing pulses are both important to the outcome.
Unless the detailed shape of white dwarf internal chemical profiles can be
measured empirically through asteroseismology, or until some independent
constraints on internal mixing are possible, this systematic uncertainty
will remain.

Switching to the same convective prescription as was used in the
spectroscopic analysis of \citet{bea99} did not resolve the systematic
errors in the derived mass and temperature of either object. This outcome
appears to be related to the simplified treatment of the surface helium
layer and the use of static envelopes. 

Models that include time-dependent
diffusion in the envelopes, such as those of \citet{dk95}, reveal that the
surface helium layers of DBV stars may contain a double-layered
structure---with a relatively thin pure He layer overlying a thicker mixed
He/C/O layer above the C/O core. The most extensive observational test for
this structure has, so far, come from \citet{fb02} who calculated a grid
of models with masses and temperatures near the spectroscopic values for
GD~358 and compared the observed and calculated periods. By including 
these double-layered envelopes---but neglecting the theoretically expected 
composition and structure in the core---the authors were able to match the 
pulsation periods of GD~358 with the same level of precision as the fit of
MWC, which used single-layered envelopes, ML3 convection, and a 
physically motivated C/O core. 

An analytical explanation of how these two models---one with extra structure 
in the core, and the other with extra structure in the envelope---can produce 
fits of comparable quality was recently provided by \citet{mon03}. 
Composition transition zones at certain locations in the core and envelope
can produce deviations from uniform period spacing that are difficult to 
distinguish from each other. This leads to a potential ambiguity between,
for example, the expected C/O transition near $0.5~m/M_*$ and the outer
He transition near $10^{-6}~m/M_*$. As a consequence, it is crucial that
we determine which feature has the largest imprint on the observed periods.

The new fit for GD~358 in Table \ref{tab1}
is more directly comparable to \citeauthor{fb02}'s fit, since it also
uses only the observed periods to determine the optimal model. Even after
accounting for the additional free parameter, the new fit has significantly
lower residuals than the pure C double-layered envelope fit, suggesting 
that the internal C/O profile may be the more important of the two
internal structures. However, the continued presence of a local minimum
with $\log(M_{\rm He}/M_*)\sim -6.0$ in fits using single-layered models 
suggests that double-layered envelopes may improve the fit even further.

\citeauthor{fb02} have also concluded that improved fits to GD~358 may 
be possible using models that include {\it both} double-layered envelopes 
and a variable core composition. However, their calculations were 
limited to model grids using several {\it uniform} C/O mixtures, so they 
did not have access to any information contained in the residuals that 
might allow a determination of the C/O profile. In the initial results 
from an ongoing study (Metcalfe et al., in preparation) we find that 
double-layered envelope models with an adjustable C/O profile yield 
significantly improved fits relative to models with a pure C core. 

Clearly, the way forward is a thorough exploration of hybrid models with 
self-consistent double-layered envelopes and physically-motivated internal 
C/O profiles. With some luck, this may eventually reduce the rms residuals 
down to the level of the observational noise, and provide the most accurate 
value possible for the \cago reaction rate.

\acknowledgements
I would like to thank Ed Nather, Don Winget, Paul Charbonneau, Maurizio
Salaris, and Gerald Handler for fruitful collaborations during the various
stages of this project.

\end{document}